\documentclass[acus]{JAC2000}

\usepackage{graphicx}
\begin{document}

\title{\flushright{WEAT002}\\[15pt]
    \centering RE-ENGINEERING OF THE GSI CONTROL SYSTEM}
\author{U. Krause, V. RW Schaa, GSI, Darmstadt, Germany}

\maketitle

\begin{abstract}
After more than 12 years of operation without substantial revision
a modernization of the control system at GSI is overdue.
A strategy to adapt the system to future needs is outlined.
The system has to support a specific environment
of which the main features are described.
More  flexibility than in the current system can be achieved
while still using many parts of the actual system.
\end{abstract}

\section{INTRODUCTION}

The actual GSI control system started operation in 1989.
Many extensions and refinements have been developed since
but no substantial revision could be made.
As a result the system is outdated in many aspects.
Only one environment is supported,
one fieldbus, one type of device controller,
one operating system for the applications.
Hardware is no longer available, e.g. the controller boards from 1990,
support for Pascal as programming language for the device control software
ended.

Most components were developed to inhouse standards.
Interfaces between components are too complex.
Therefore exchange of single components is costly.
A general revision is overdue.
The modernization has to consider the characteristics of the GSI environment.

\section{CONTROL ENVIRONMENT}

\subsection{GSI Accelerator Operation}

GSI operates three accelerators for all kind of ions from hydrogen
to uranium: The linear accelerator Unilac, the synchrotron SIS and
the storage ring ESR. Linac and synchrotron are operated in a
pulse to pulse time sharing mode with a repetition rate of 50~Hz
of the linac and 0.1 to 0.5~Hz of the synchrotron. Switching to
different ion species, energies, and experimental targets is done
with this rate. Three independent ion sources serve in parallel typically
five experiments at Unilac, SIS and ESR. The average duration of an
experiment is one week.

In addition to this flexible experimental operation a rigid mode
for heavy ion cancer therapy is provided \cite{med}:
Any carbon beam from a fixed set of 254 energies,
15 intensities, and 7 spot sizes will be delivered
to the irradiation place by request.

\subsection{Device Handling}

Pulse to pulse switching demands a more complex device handling
than simple schemes like `set reference', `read actual value'.
This can be illustrated with magnets
in Unilac cycles for low-charged ions.
One controller has to service up to 12 magnets in 20~ms cycles.
Preset time has to be maximized to reach high currents,
and stable current duration has to be limited to reduce thermal load.

Broadcasting a medium level set value at start of cycle
gives slow devices time to reach full current.
To avoid overheating of fast devices the dedicated
set value is delayed for 4~ms. After end of beam all
magnets have to be set to a zero value. The duration of stable
currents is too short to read actual values directly. So `sample
and holds' are triggered for read-out. The controller can
read these values not until in the following cycle
reference values have been set.

\subsection{Device Realization}

In most cases the devices connected to the control system are rather complex.
Some components are implemented as distinct parts
like the extraction kicker with 28 modules.
On the other hand some hardware components host several devices,
e.g. profile grids.
Up to 16 grids are distributed on 8 channels in one measuring device.
Multiplexing restricts measurement to one channel per cycle.
Nevertheless the variety of different implementations
has to be presented in a comprehensive way to the operation crew.
Grids have to appear as independent devices
and the kicker modules have to be combined to one kicker.

\section{ACTUAL CONTROL SYSTEM}
\subsection{Scheme of the control system}

\begin{figure}[h]
 \centering
 \includegraphics*[width=75mm]{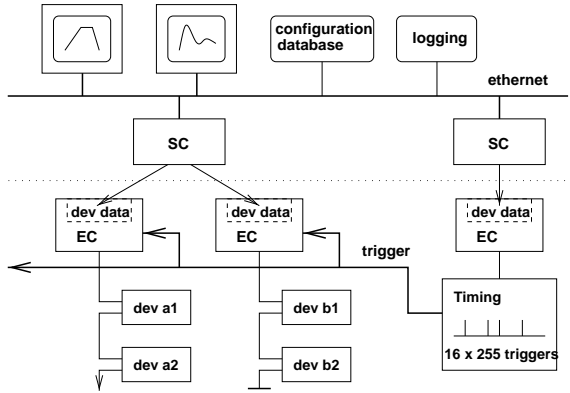}
 \caption{Schematic view of the actual control system\label{ctractarch}}
\end{figure}

The hardware outline of the actual control system \cite{siscos} is given
in figure \ref{ctractarch}.
The diagram reflects also the logical view
since each module is rigidly connected to one of the hardware levels.

All devices are linked via field bus (modified MIL 1553)
to distributed equipment controllers (EC).
Synchronized operation is achieved by triggers from programmable
central timing units, one for each accelerator.
Supervisory controllers (SC) handle interaction with the operation level.
One SC serves up to nine ECs.

Communication between EC and SC is done by dual ported RAM on the EC.
SCs and workstations communicate via ethernet
by an in-house protocol comparable to UDP/IP.
VME boards with 68020 processors are used for ECs and SCs.
Operation level workstations run OpenVMS.

\subsection{Real-Time Control}

The real-time level is the most substantial part of the GSI control system.
Device interactions are triggered by signals from central timing units.
Up to 255 different triggers allow flexible adaption
to the accelerating process.

ECs run autonomously under control of the timing unit
after device data have been supplied.
Up to 16 different sets of data, called virtual accelerators, may
be configured in parallel on the ECs.
This enables pulse to pulse switching between as many different beams.
Beams for cancer therapy are handled analogously \cite{med}.

The sequence of virtual accelerators
is determined online by the timing units:
Beam is produced only on request by the experimental area.

Execution of commands from the operation level is provided.
Device interrupts and polling services for survey of the devices
are supported.

\subsection{Device Representation}

The devices are represented in an object oriented manner
as independent units even though the control system was developed
in a procedural way.
Unique device names, the so-called nomenclatures, facilitate addressing.
Every property is modelled by an action,
coded as a procedure on the SC,
with data to be exchanged,
e.g. sending a reference or reading an actual value.
Properties are identified by name and described in a formalized way
by type and count of corresponding data.
Based on this description one single interface allows
access to every property of every device.

To keep applications well structured nomenclatures represent
independent objects with relevance for the process of acceleration.
These logical devices are constituted on the real-time level:
Every nomenclature must have a corresponding entry on the EC.
Mapping of the connected hardware to the operations view is demanded.

\section{UPGRADES}
\subsection{Strategy}

Replacement of the control system by a different one would require a lot of
effort.
Existing devices and interfaces must be supported,
the achieved quality of accelerator operation has to be provided,
and existing control hardware has to be used further
in order to reduce expenses.

Actually 2750 devices are controlled by 256 ECs in 41 VME crates.
Device specific adaptions can be subsumed in 61 classes,
each requiring its own handling.
Implementation of device adaptions requires 145,000 lines of code (LOC)
compared to 26,000 LOC for common system software.
The effort needed to reimplement peculiarities could
be seen when a functional prototype of the Unilac timing unit was rebuild.
Although it provided less functionality than is implemented today,
and work was done within the same environment,
it took about one person year.

Fortunately the architecture of the existing control system
is still up to date.
Stepwise migration to a system similar to the actual one but
providing greater flexibility is managable.
This will result in a to date system
and will allow to re-use most of spent investments.

\subsection{Outline of the Future Control System}

\begin{figure}[h]
 \centering
 \includegraphics*[width=80mm]{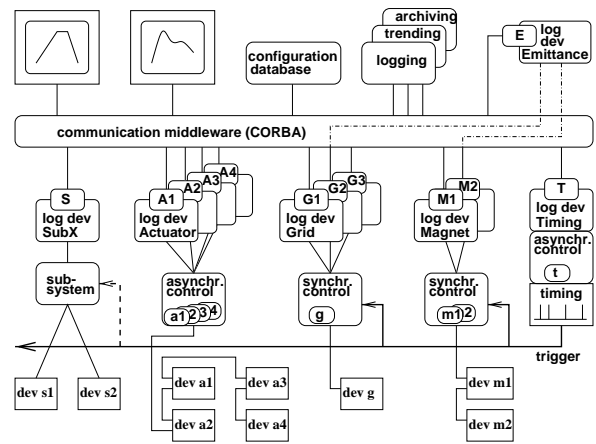}
 \caption{Schematic view of the future control system\label{ctrfutarch}}
\end{figure}

The outline of the proposed future control system is shown
in figure \ref{ctrfutarch}.
It is still a three level approach: Device control engines,
device representation by logical devices, and the application level.
Different from figure \ref{ctractarch} it is a logical view.
Components can be installed on every hardware level of the control system.

\subsection{Synchronized Device Control}

Strong point in the existing control system is the real-time device control.
It proved to be very well adapted to the needs of
flexible pulse to pulse operation of the GSI accelerators.
Therefore the same mechanism will be used in the future.
This allows to utilize the existing generation and delivery of
triggers for device synchronization.

Actions to service devices autonomously, under control of the timing system,
are realized in synchronous device control engines.
The actual EC software can serve as a first version of these engines.
Fortunately most effort to implement device specific adaptions was on the EC,
both in size and complexity.

\subsection{Extensions for the Device Control Level}

In many cases a synchronization with the accelerating cycle is not needed,
e.g. for slow mechanical actuators.
An asynchronous device control engine will be provided
by reducing a synchronous engine to polling and
service of device interrupts.

Real-time systems are difficult to survey and to debug.
To keep the control engines simple
any mapping between the hardware and the process view should be avoided here.
E.g. the electronics for profile grids should be handled as one unit.
Consequent reduction to the kernel functionality allows
in many cases to describe device characteristics by simple tables.
These devices then can be handled by one common software.

In the current EC software,
the hardware dependent parts of code have to be identified
to make the control engines portable.
Implementation will be possible on any computer in the control system,
on dedicated device controllers and on multi purpose computers.
Nevertheless if good real-time performance is needed
computers with a hard real-time system will be used.

\subsection{Device Modelling Level}

Object oriented approach suggests to represent connected hardware
as objects, called logical devices. Device functionality in the
sense of the actual control system then corresponds to methods of
the logical device.

Dealing with various device specific methods can be confusing.
Therefore every logical device will have a common method as interface.
Variable data formats can be exchanged
by using types like CORBA's \texttt{any}.
In IDL notation this method may look in principle like \newline
\hspace*{4mm}\texttt{\small void access(in string property, inout any data)}\\*[0mm]
\hspace*{10mm}\texttt{\small raises (ControlException)};

Logical devices are suitable locations to map the accelerator hardware
to logical devices oriented to the process.
Different abstraction levels can be build by cascading.
E.g. magnets and profile grids can be combined
to emittance measurement devices.
Any unit can be integrated as a logical device at this level:
Complete subsystems like SCADA systems,
devices with an OPC interface,
or even software objects like databases.

A scheme for the logical devices has to be developed
which is more general than in the actual system.
Effort for implementation of device specific adaptions
can be reduced by integrating code from the existing system.
Procedures corresponding to properties can be transformed to
objects' methods.
Header and exit part of the procedures have to be replaced
but the body can be kept with slight modifications.
This allows a fast integration of the existing implementations
but of course can only be a first step.
Modern software development technics allow much more elegant
solutions compared to realisations in the actual system.

\subsection{Networking}

Communication with logical devices, distributed objects,
is straightforward.
Common middleware like COR\-BA based systems support
all needs of accelerator operation.
Naming services or alternatively explicit handling of object references
allow addressing of logical devices by name.

\subsection{Application Level}

No detailed investigations have been made yet.
OpenVMS and Unix are based on similar concepts.
This suggest to use Linux as future basis for the application level.

In a transitional period both operation systems have to be supported
in parallel.
To allow further usage of existing applications the current interface
for device access has to be provided in the future too.

\subsection{Preparatory Work}

The device control software was written in Pascal.
Actual software development systems are now based on C++.
To enable future usage of existing software conversion of the code
to C has started \cite{p2c}.

Substantial changes of the actual system can easily impact
the ongoing accelerator operation.
To limit the implications of modifications the modularization
has to be enhanced.
In a re-engineering process each module in the control system has to be
provided with structured interfaces to reduce coupling. Only after this
it will be possible to replace existing components or port
components to other platforms with acceptable effort.

\section{CONCLUSION}

The paper outlines a strategy for a rejuvenation of the existing
control system. It shows the possibility to enhance flexibility and
capacity of the system and nevertheless to integrate many parts of
the existing system.
The modernized control system will be suitable
for the proposed new accelerator facilities too.


\begin{thebibliography}{9}

\bibitem{med}
U. Krause, R. Steiner,
``Adaption of a Synchrotron Control System for Heavy Ion Tumor Therapy'',
Proceedings of ICALEPCS '95, Chicago, USA, 1995

\bibitem{siscos}
U. Krause, V. Schaa, R. Steiner, ``The GSI Control System'', Proceedings
of ICALEPCS '91, Tsukuba, Japan, 1991.

\bibitem{p2c}
L. Hechler,
``Converting Equipment Control Software from Pascal to C/C++'',
these proceedings.

\end{thebibliography}
\end{document}